\documentclass{article}

\usepackage{arxiv}
\usepackage[utf8]{inputenc} 
\usepackage[T1]{fontenc}    
\usepackage{hyperref}       
\usepackage{url}            
\usepackage{booktabs}       
\usepackage{amsfonts}       
\usepackage{nicefrac}       
\usepackage{microtype}      
\usepackage{lipsum}
\usepackage{graphicx}
\graphicspath{ {./images/} }
\title{DYNAMIC ACCOMMODATION MEASUREMENT USING PURKINJE IMAGES AND ML ALGORITHMS}
\author{
Faik Ozan Ozhan \\
  Koc University\\
  Istanbul, Turkey, 34450 \\
  \texttt{fozhan21@ku.edu.tr} \\ \And
 Arda Gulersoy \\
  Koc University\\
  Istanbul, Turkey, 34450 \\
  \texttt{agulersoy19@ku.edu.tr} \\ \And
  Ugur Aygun \\
  Koc University\\
  Istanbul, Turkey, 34450 \\
  \texttt{uaygun@ku.edu.tr} \\ \And
Afsun Sahin \\
Koc University\\
Istanbul, Turkey, 34450\\
\texttt{afsahin @ku.edu.tr} \\ \And
Hakan Urey\\
Koc University\\
Istanbul, Turkey, 34450 \\
\texttt{hurey@ku.edu.tr} \\
}

\begin{document}

\maketitle

\begin{abstract}

We developed a prototype device for dynamic gaze and accommodation measurements based on 4 Purkinje reflections (PR) suitable for use in AR and ophthalmology applications. PR1\&2 and PR3\&4 are used for accurate gaze and accommodation measurements, respectively. Our eye model was developed in ZEMAX and matches the experiments well. Our model predicts the accommodation from 4 diopters to 1 diopter with better than 0.25D accuracy. We performed repeatability tests and obtained accurate gaze and accommodation estimations from subjects. We are generating a large synthetic data set using physically accurate models and machine learning.

\end{abstract}


\section{Introduction}
As eyes look at different points in space, some processes are going on in them. The effective focal length of the eyes is
adjusted by a change in the shape of the human crystalline lens controlled by ciliary muscles. This process is called
accommodation and the distance where the eyes are focused is called accommodation depth\cite{chakravarthula2018focusar}.
\begin{figure}
    \centering
    \includegraphics[width=0.65\linewidth]{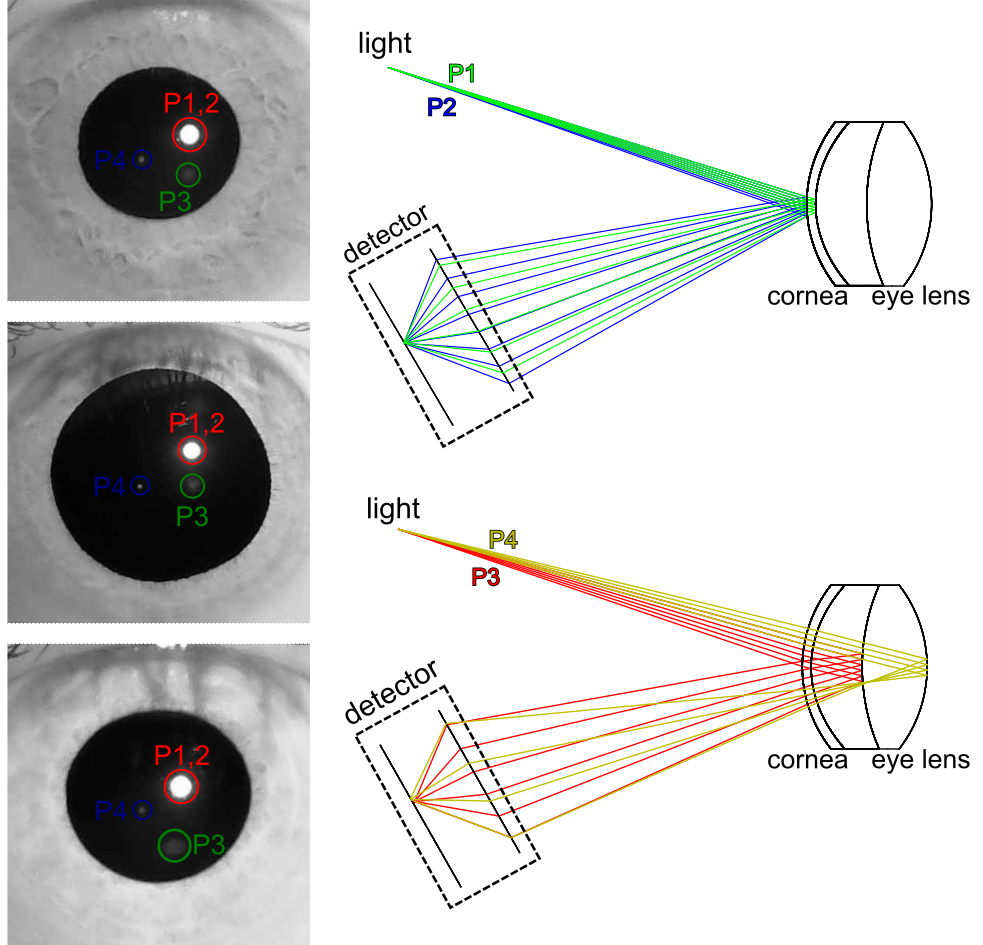}
    \caption{(a) Captured image illustrating Purkinje reflections, (b) ZEMAX model illustrating imaging setup and Purkinje reflections.}
    \label{fig1}
\end{figure}

In addition, eyes converge to fixate upon points in space. The distance between a specified point and the eyes is called vergence distance. Accommodation and vergence responses follow each other. Even if they follow each other, there are some mismatches in some cases. This is called vergence-accommodation conflict (VAC)\cite{geng2013three}. So, even if accommodation is somehow predicted by vergence information, there are mismatch cases like 3D films and monocular applications. In
ophthalmology, eye accommodation depth is measured for various applications such as the diagnosis of presbyopia and anisometropic amblyopia.

In this work, we developed a prototype device for dynamic gaze and accommodation measurements based on 4 Purkinje reflections (PR) suitable for use in AR and ophthalmology applications. PR1\&2 and PR3\&4 are used for accurate gaze and
accommodation measurements, respectively.

\section{Method}
\subsection{Purkinje Images: Their Formation, Relation to Accommodation and Vergence}
Purkinje images (or reflections) are reflections from various layers of the human eye. Four of these Purkinje reflections can
be captured and specified clearly. There are named P1, P2, P3, and P4. P1 is the reflection from the anterior and P2 from
the posterior surface of the cornea, P3 from the anterior, and P4 from the posterior surface of the human crystalline lens.
The formation of these images can be understood well by modeling the eye and observing the reflections in optical design
software illustrated in Figure \ref{fig1}.

\begin{figure}
    \centering
    \includegraphics[width=0.85\linewidth]{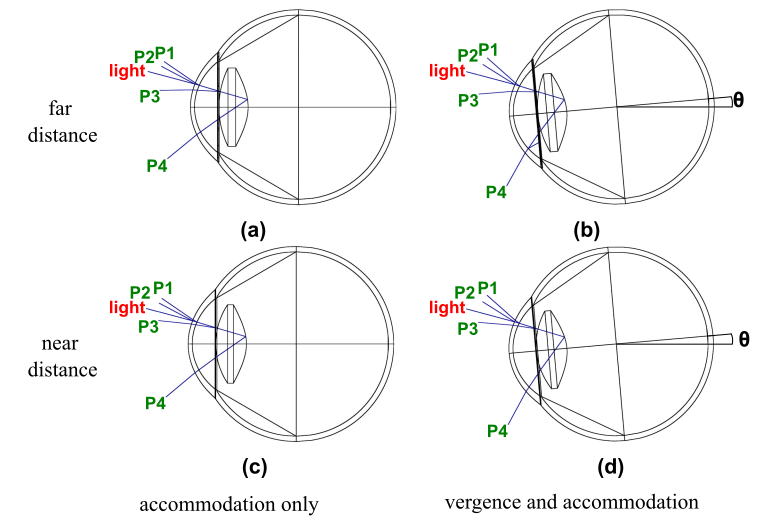}
    \caption{ZEMAX simulation layouts illustrating change of locations of Purkinje chief ray reflections with (a,b) accommodation (focus change) and (c,d) vergence (eye rotation)}
    \label{fig2}
\end{figure}

As the eye accommodates (its focus change), the ciliary muscles contract, and the curvature of the lens increases changing
the location of Purkinje reflections. As the eye converges to a point (it rotates), its shape in 3D changes resulting in a change
of locations of reflections. In Figure \ref{fig2}, the change of locations of Purkinje images with accommodation and vergence is
shown based on our eye model.
\subsection{Experimental System Design}
\begin{figure}[ht]
    \centering
    \includegraphics[width=\linewidth]{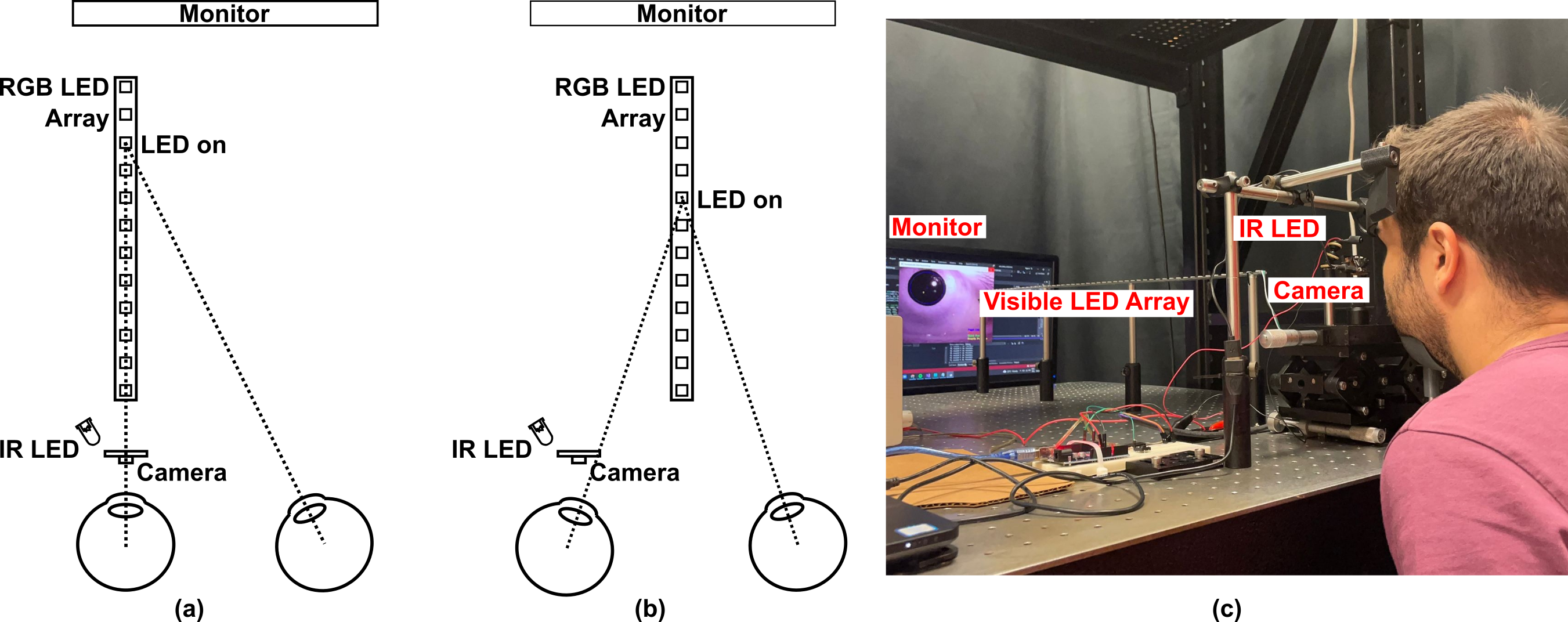}
    \caption{Setup schematic designed for measuring (a) accommodation effect, (b) accommodation and vergence effects, (c) picture of setup}
    \label{fig3}
\end{figure}

We designed a setup that enables us to relate Purkinje images with both accommodation and vergence. For this purpose, we
used an RGB LED array placed in the visual field of the eye as target points. We turned on and off some of these LEDs on
the LED array sequentially and asked for the user to accommodate their eyes to the light source that is turned on during the
experiments. Our experimental setup schematics which are designed for measuring effects of both vergence and
accommodation and only effect of accommodation is illustrated in Figure \ref{fig3}.
\section{Results}
\subsection{Detection of Purkinje Reflections}
\begin{figure}[ht]
    \centering
    \includegraphics[width=\linewidth]{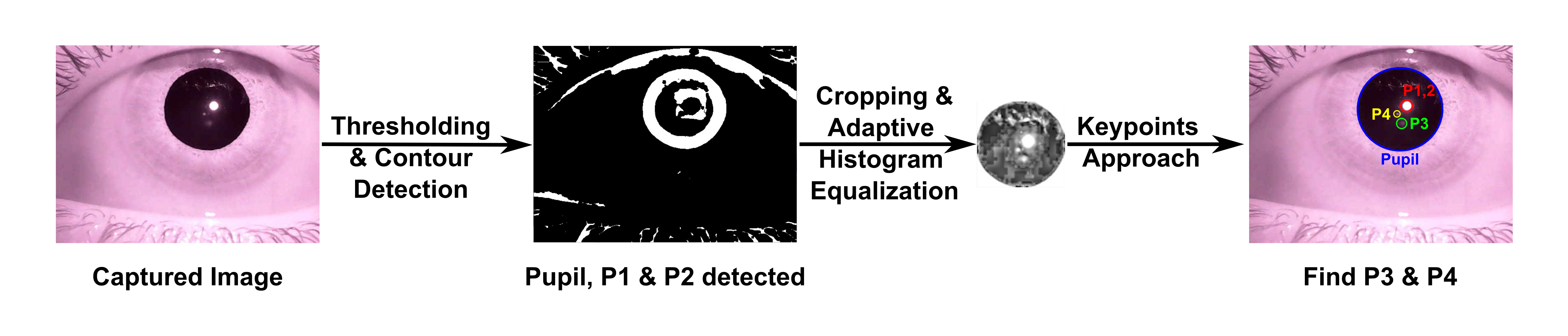}
    \caption{Purkinje reflections detection algorithm pipeline}
    \label{fig4}
\end{figure}

To determine the relationship between the locations of Purkinje reflections and generate labeled data, we first needed to
find the locations of Purkinje reflections on captured images from the camera by various image processing algorithms. Our
algorithm to find the locations of these images are summarized in Figure \ref{fig4}.

\begin{figure}[ht]
    \centering
    \includegraphics[width=\linewidth]{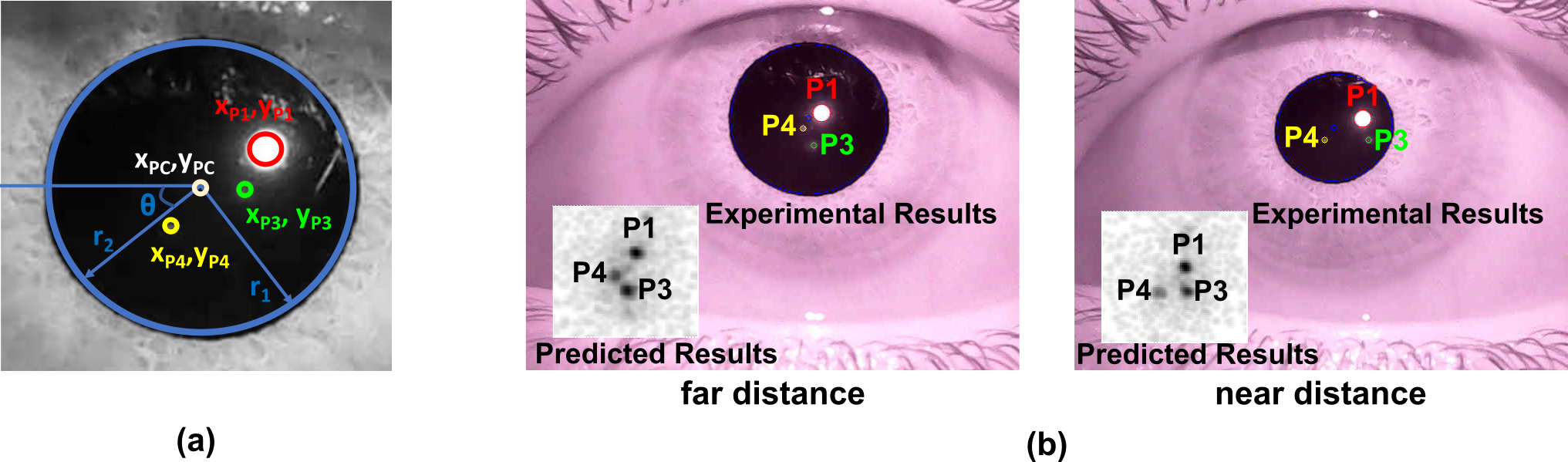}
    \caption{(a) All ten parameters, (b) Image processing results and ZEMAX Predictions match.}
    \label{fig5}
\end{figure}

Purkinje reflections are found using image processing using the captured images on the camera. ZEMAX non-sequential
ray tracing simulations resulted in a similar change in Purkinje reflection locations in response to changes in
accommodation and vergence as illustrated in Figure 5. Using the results, we identified ten parameters, which are pupil center and size ($x_{PC}$, $y_{PC}$, $r_{1}$,$r_{2}$) and locations of Purkinje images ($x_{P1}$, $y_{P1}$, $x_{P3}$, $y_{P3}$, $x_{P4}$, $y_{P4}$). All of those parameters seen in Figure \ref{fig5}.
\subsection{Results from Setup designed for measuring both accommodation and vergence effect}
\begin{figure}[ht]
    \centering
    \includegraphics[width=0.45\linewidth]{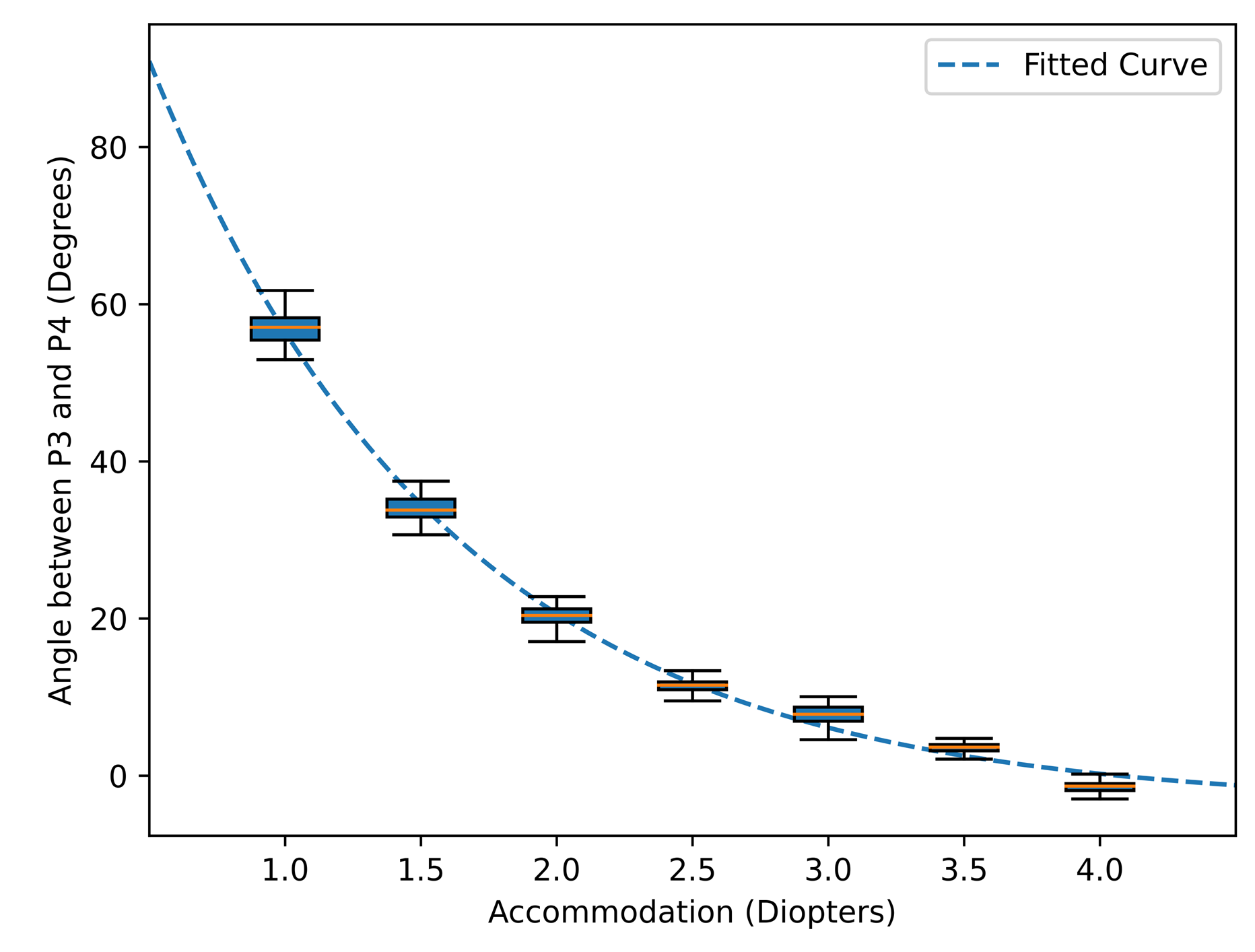}
    \caption{Accommodation versus angle between P3\&P4 measured with one subject. Error bars indicate extremities and quartile range using 300 frames collected in 6 seconds after allowing for 2-3 seconds for transition.}
    \label{fig6}
\end{figure}

When both vergence and accommodation effects are present, we figured out that the IR LED location is critical to see all
Purkinje images. For most of the cases, we saw that at least one of the Purkinje images is blocked by the iris, and not visible.
We performed ZEMAX analysis and swept the LED source on a grid to see which LED-camera configuration is optimum
and all Purkinje reflections are visible in the 1 to 4 diopters range. The LED location of the experimental setup is arranged
according to simulation results. When the results are analyzed, it can be seen that the angle between the third and fourth
Purkinje images changes in correlation with accommodation and vergence. Furthermore, it can be said that the angle
changes almost exponentially when both accommodation and vergence effects are present. These results are shown in Figure \ref{fig6}.

\subsection{Results from Setup designed for measuring only accommodation effect}
\begin{figure}[ht]
    \centering
    \includegraphics[width=0.9\linewidth]{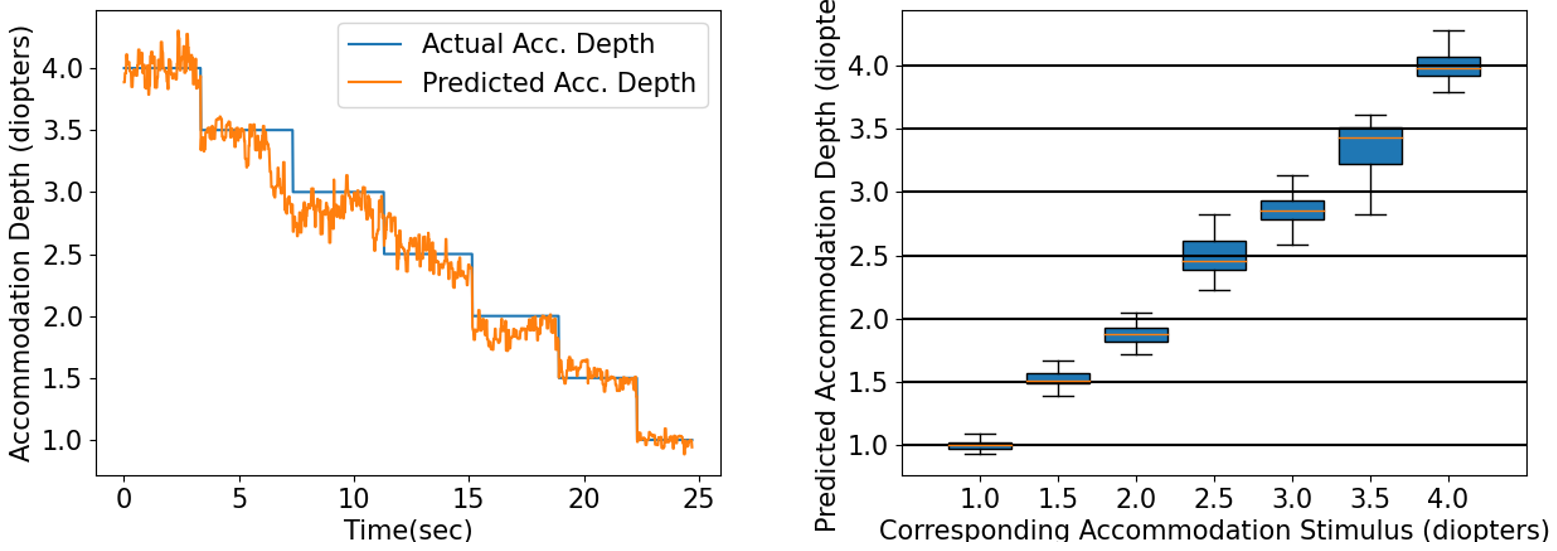}
    \caption{(a) Change of predicted accommodation depth and corresponding accommodation stimulus with time, (b) Distribution of predicted accommodation values at corresponding accommodation stimulus. Error bars indicate extremities and quartile range using 300 frames collected in 6 seconds after allowing for 2-3 seconds for transition.}
    \label{fig7}
\end{figure}

When the accommodation effect is present, the angle between P3 and P4 can be calculated as in the previous case. However,
the exponential correlation isn’t observed for the accommodation case. Moreover, accommodative response fluctuates, so
angle-based measurement doesn’t give accurate results. We propose that machine learning (ML) algorithms give more accurate results for accommodation measurement. The accommodation values for one subject using our ML algorithm are
shown in Figure \ref{fig7}.
\section{Conclusion}
In this study, we proposed a prototype device for dynamic gaze and accommodation measurements based on 4 Purkinje
reflections (PR). Our model predicted the accommodation from 4 diopters to 1 diopter with RMSE of lower than 0.2D for
subjects. The results show that the proposed device is promising especially in measuring accommodation depth and
vergence in ophthalmic applications and near-eye displays.
\section{Funding}
This project is sponsored by the European Innovation Council’s HORIZON-EIC-2021-TRANSITIONCHALLENGES Program, Grant Number 101057672 and Tübitak’s 2247-A National Lead Researchers Program,
Project Number 120C145.

\bibliographystyle{unsrt}  
\bibliography{references} 

\end{document}